\documentclass[%
 reprint,
 amsmath,amssymb,
 aps,
prl]{revtex4-2}
\usepackage{graphicx}% Include figure files
\usepackage{float}%use [H] when adding graphics so that they are in the exact position as in the code
\usepackage{dcolumn}% Align table columns on decimal point
\usepackage{makecell}

\usepackage{amsmath}%for formula formatting
\usepackage{amssymb}

\usepackage{tensor} %do T\indices{_c_b^d_e}
\usepackage{slashed}%do \slashed p
\usepackage{mathtools}%inclusion map: \xhookrightarrow{}
\usepackage{leftidx}%Left and right subscripts and superscripts in math mode
\usepackage{esint}%Extended set of integrals for Computer Modern
\usepackage{amsfonts,amsthm,bm}%boldface math symbols
\usepackage{empheq}

\usepackage{physics}
\usepackage{feynmf}
\usepackage{braket}
\usepackage{youngtab}

\usepackage{xcolor}
\usepackage{color}
\definecolor{darkgreen}{rgb}{0,0.5,0}
\definecolor{darkblue}{rgb}{0,0,0.6}
\definecolor{purple}{rgb}{0.4,.2,0.7}
\usepackage{colortbl}      % Allows cell-specific coloring in tables

\newcommand{\nn}{\nonumber}

\begin{document}
\preprint{APS/123-QED}

\title{Horizon Edge Partition Functions in $\Lambda>0$ Quantum Gravity}% Force line breaks with \\

\author{Y.T. Albert Law}
 \email{ytalaw@stanford.edu}
\affiliation{%
 Leinweber Institute for Theoretical Physics at Stanford, 382 Via Pueblo, Stanford, CA 94305, USA
}%
\author{Varun Lochab}%
 \email{vv2338@columbia.edu}
\affiliation{%
 Center for Theoretical Physics, Columbia University, New York, NY 10027, USA
}%

\date{\today}% It is always \today, today,

\begin{abstract}

We obtain the spectra of codimension-2 horizon ``edge” degrees of freedom for gravity and higher-spin gauge fields in de Sitter space and in the static Nariai spacetime, advancing previous Lorentzian and Euclidean analyses of one-loop thermodynamics. The edge spectra exhibit universal shift symmetries, revealing a novel symmetry-breaking structure in one-loop partition functions with positive cosmological constant. For the graviton, these modes admit a geometric interpretation as fluctuations of the cosmic horizon, which also persists in the Nariai case.

\end{abstract}

\maketitle

\section{\label{sec:intro}Introduction}

An outstanding challenge in theoretical physics is to construct a UV-complete framework for quantum gravity in de Sitter (dS)-like spacetimes, which approximate our universe at very early \cite{Starobinsky:1979ty,Guth:1980zm,Linde:1981mu,Albrecht:1982wi} and very late \cite{SupernovaCosmologyProject:1996grv,SupernovaSearchTeam:1998fmf,SupernovaCosmologyProject:2008ojh,Loeb:2001dh,Krauss:2007nt} times. Pending a fully explicit top-down construction, a natural bottom-up strategy is to study quantities in the low-energy effective theory that are invariant under diffeomorphisms and field redefinitions. In the present context, one such quantity is the Euclidean gravitational path integral with $\Lambda>0$ \cite{Gibbons:1976ue}
\begin{align}\label{introeq:Z}
    \mathcal{Z} = \int \mathcal{D}g \, \mathcal{D}\Phi\, e^{-S[g,\Phi]} \;,
\end{align}
where $g$ denotes the metric and $\Phi$ other fields. In saddle-point approximation, the sum runs over closed Euclidean manifolds of positive curvature. This object provides a rare infrared probe of microscopic models, much as in black hole thermodynamics or AdS/CFT. 
%\VL{Sen and all papers}. 
Unlike those cases, however, without an asymptotic boundary from which to extract diffeomorphism-invariant correlators,
%\VL{Witten and all papers with Witten diagram etc}\AL{Not necessary. That's too remote to be cited here.}
 \eqref{introeq:Z} appears at first sight as a featureless number, obscuring its meaning. One natural interpretation is that $\log \mathcal{Z}$ represents a microcanonical entropy, as explored in earlier work \cite{Banks:2000fe,Banks:2003cg,Witten:2001kn} and more recent developments \cite{Anninos:2021ene,Coleman:2021nor,Anninos:2021eit,Anninos:2022ujl,Bobev:2022lcc,Collier:2025lux}. To make further progress, additional guiding principles would be valuable for refining \eqref{introeq:Z} into an interpretable form.

\begin{figure}[ht]
	\centering
	\includegraphics[width=0.45\textwidth]{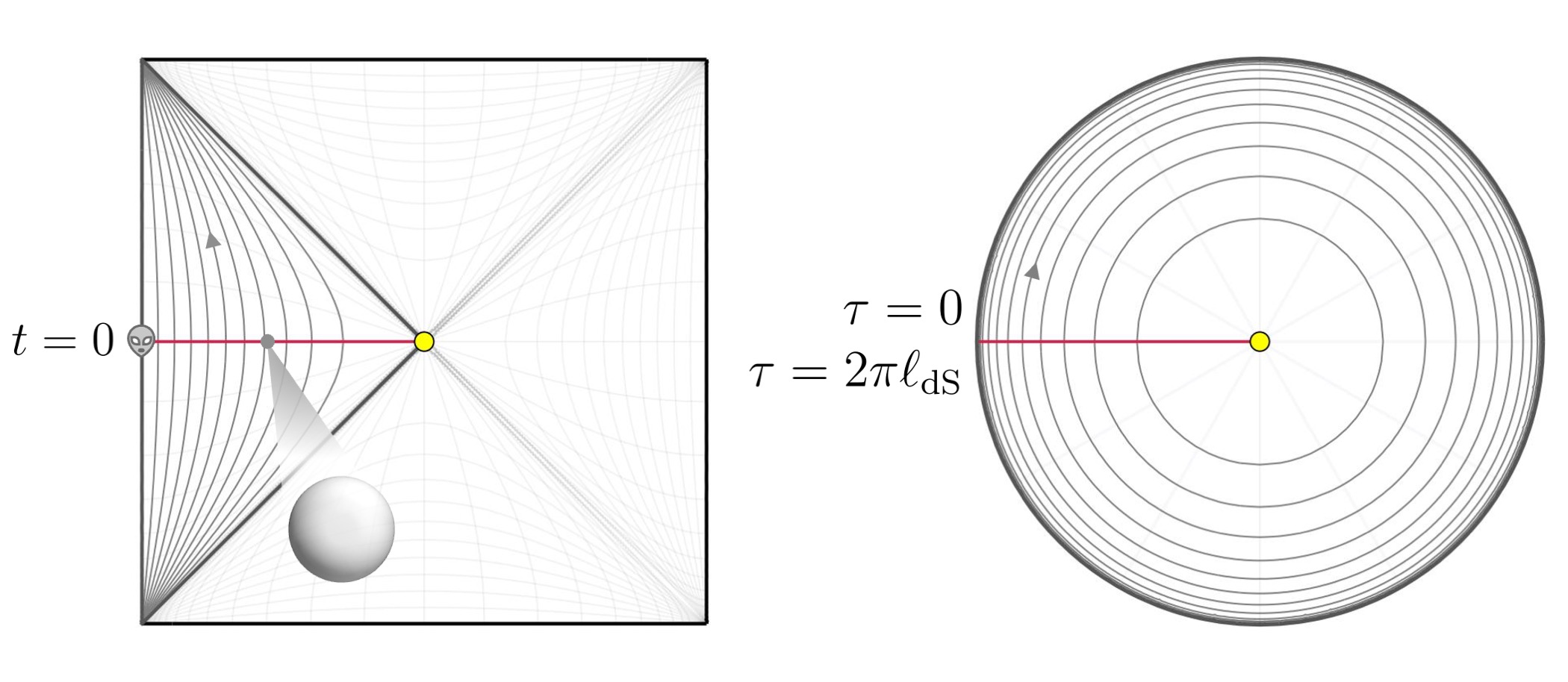}
	\caption{A round $S^{d+1}$ arises from a $dS_{d+1}$ static patch by Wick-rotating the observer's proper time and imposing periodicity, $t \to -i\tau$ with $\tau \sim \tau + 2\pi\ell_\text{dS}$. { This suggests interpreting the $S^{d+1}$ path integral as a thermal trace, analogous to Euclidean methods in quantum field theory at finite temperature. Subtleties may arise, however, at the (Euclidean) horizon or ``bolt'' \cite{Gibbons:1979xm}, the codimension-2 fixed point set (yellow dot) of Euclidean time evolution where the thermal cycle degenerates.}  
    %The $S^{d+1}$ path integral is thus naturally interpreted as a thermal trace, though subtleties may arise at the (Euclidean) horizon or ``bolt'' \cite{Gibbons:1979xm}, the codimension-2 fixed point set (yellow dot) of Euclidean time evolution.
    }
	\label{pic:dSSP}
\end{figure}

Quantum corrections about the round $S^{d+1}$ saddle of \eqref{introeq:Z}, interpreted thermodynamically in a Lorentzian $dS_{d+1}$ static patch (Fig.~\ref{pic:dSSP}) \cite{Figari:1975km,PhysRevD.15.2738}, reveal unexpectedly rich structure. 
At one loop, the $S^{d+1}$ partition function for fields of arbitrary mass and integer spin universally factorizes as \cite{Anninos:2020hfj}\footnote{{ Our convention for $Z_{\rm edge}$ differs from \cite{Anninos:2020hfj} by inversion. We adopt the present notation because recent Lorentzian analyses naturally interpret $Z_{\rm edge}$ itself as a partition function, at least for $p$-form gauge theories \cite{Ball:2024hqe,Ball:2024xhf}.}}
\begin{align}\label{introeq:PIsplit}
	Z^\text{1-loop}_\text{PI}\left[S^{d+1}\right]
	= Z_{\rm bulk}(\beta=\beta_\text{dS})\, Z_{\rm edge}.
\end{align}
The bulk factor $Z_{\rm bulk}$ is the thermal partition function of an ideal bosonic gas in a $dS_{d+1}$ static patch at inverse dS temperature $\beta_{\rm dS}=2\pi \ell_{\rm dS}$ with $\ell_\text{dS}=\sqrt{\tfrac{d(d-1)}{2\Lambda}}$, 
\begin{align}\label{introeq:Zbulk}
    \log Z_{\rm bulk}(\beta) = \int_0^\infty \frac{dt}{2t}\,
    \frac{1+e^{-\tfrac{2\pi}{\beta}t}}{1-e^{-\tfrac{2\pi}{\beta}t}}\,\chi(t)\;.
\end{align}
Here $\chi(t)$ is the $SO(1,1)$ character of the boost generator in a given $SO(1,d+1)$ unitary irreducible representation (UIR), encoding the spectrum of quasinormal modes (QNMs) \cite{Sun:2020sgn},
\begin{align}\label{introeq:charQNM}
    \chi(t) = \sum_{\text{QNM}} N_z\, e^{-iz|t|}\;,
\end{align}
with $z$ the QNM frequency and $N_z$ its degeneracy. For scalars, $\chi(t)$ can also be obtained as a spatially integrated correlator \cite{Grewal:2024emf}.

Our focus will be on the second factor in \eqref{introeq:PIsplit}, $Z_{\rm edge}$, which arises only for spin $s \geq 1$. It is termed ``edge'' because it takes the form of an {\it inverse} path integral on $S^{d-1}$ (radius $\ell_\text{dS}$), naturally associated with the cosmic horizon of a $dS_{d+1}$ static patch. Table \ref{tab:table1} summarizes $1/Z_\text{edge}$ for several $p$-form theories. In the Appendix, we summarize the explicit formulas \eqref{introeq:PIsplit} and \eqref{introeq:Zbulk} specialized to gravity.
\begin{table}[H]
\caption{\label{tab:table1}%
$1/Z_\text{edge}$ for Maxwell \cite{RiosFukelman:2023mgq,Ball:2024hqe}, Yang–Mills (1-loop) \cite{Law:2025ktz}, and $p$-forms \cite{Mukherjee:2023ihb,Ball:2024xhf}.
}
\begin{ruledtabular}
\begin{tabular}{cc}
Theory on $S^{d+1}$ & \textrm{$1/Z_\text{edge}$} \\
\colrule
$U(1)$ Maxwell & $U(1)$ scalar \\
Yang–Mills with gauge group $G$ & $G$-valued scalars \\
$U(1)$ $p$-form & $U(1)$ $(p\!-\!1)$-form \\
Massive $p$-form & Massive $(p\!-\!1)$-form \\
\end{tabular}
\end{ruledtabular}
\end{table}

The structure \eqref{introeq:PIsplit} has well-known precedents in Maxwell theory. Early studies of entanglement entropy revealed a negative contact term \cite{Kabat:1995eq}. Later, in calculations of 4D Maxwell entanglement entropy across a ball in flat space, one maps the problem to a $dS_4$ static patch and finds that a thermal result \cite{Dowker:2010bu,Eling:2013aqa} differs from the conformal-anomaly expectation \cite{Solodukhin:2008dh,Casini:2011kv}. Both the contact term and this 4D discrepancy were subsequently understood in terms of ``edge modes'' localized on the entangling surface \cite{Donnelly:2011hn,Donnelly:2012st,Eling:2013aqa,Radicevic:2014kqa,Donnelly:2014gva,Donnelly:2014fua,Huang:2014pfa,Ghosh:2015iwa,Hung:2015fla,Aoki:2015bsa,Donnelly:2015hxa,Radicevic:2015sza,Pretko:2015zva,Soni:2015yga,Zuo:2016knh,Soni:2016ogt,Delcamp:2016eya,Agarwal:2016cir,Blommaert:2018rsf,Blommaert:2018oue,Freidel:2018fsk,Ball:2024hqe}. The factorization \eqref{introeq:PIsplit} may be viewed as a generalization of this phenomenon to arbitrary masses and spins in any dimension.

In this Letter, we initiate a systematic study on $Z_\text{edge}$ for symmetric tensor fields of arbitrary mass and spin. For gravity and higher-spin gauge theories, we determine their explicit edge spectra, which display universal patterns of shift symmetries. As we shall see, this structure persists for gravitons on the static Nariai geometry.

%%%%%%%%%%%%%%%%%%%%%%%%%%%%%%%%%%%%%%%%%%%%%%%%%%%%%%%%%

\section{Gravitons on $S^{d+1}$}\label{sec:gravity}

A key obstacle in formulating an edge-mode interpretation for gravity \cite{Donnelly:2016auv,Geiller:2017xad,Speranza:2017gxd,Geiller:2017whh,Freidel:2019ees,Benedetti:2019uej,Takayanagi:2019tvn,Freidel:2020xyx,Freidel:2020svx,Freidel:2020ayo,Donnelly:2020xgu,Ciambelli:2021vnn,Carrozza:2021gju,Ciambelli:2021nmv,David:2022jfd,Carrozza:2022xut,Ciambelli:2022cfr,Mertens:2022ujr,Wong:2022eiu,Donnelly:2022kfs,Ciambelli:2022vot,Balasubramanian:2023dpj,Lee:2024etc,Blommaert:2024cal,Fliss:2025kzi}, analogous to the Maxwell or $p$-form cases, is that its $Z_{\rm edge}$ receives contributions from multiple field types. Explicitly, it is not obvious how to disentangle the ``collapsed" expression obtained in \cite{Anninos:2020hfj},
\begin{align}\label{eq:Zedgeold}
    Z_{\rm edge}	& \propto \exp\bigg\{-\int_0^\infty \frac{dt}{2t}\frac{1+q}{1-q}\nn\\
   & \qquad  \times \left[(d+2)\frac{q^{d-1}+q^{-1}}{(1-q)^{d-2}}-\frac{q^{d}+q^{-2}}{(1-q)^{d-2}} \right]_+ \bigg\}\;,
\end{align}
into interpretable lower-spin contributions. Here $q\equiv e^{-t/\ell_\text{dS}}$ and the notation $\left[\cdots \right]_+$ is defined in \eqref{eq:flipnotation}.

\subsection{Branching the sphere}

We took on the challenge of analyzing the detailed $\mathfrak{so}(d)$ structure underlying \eqref{eq:Zedgeold}. The starting point is that one-loop $S^{d+1}$ path integrals are expressed as determinants of Laplacians on $S^{d+1}$, i.e. as infinite products over spherical harmonics furnishing irreducible representations (irreps) of $\mathfrak{so}(d+2)$:
\begin{align}\label{eq:Zprebranch}
    Z^{\text{1-loop}}_{\text{PI}}\!\left[S^{d+1}\right] \sim \prod \bigl(\mathfrak{so}(d{+}2)\text{ irreps}\bigr)\;.
\end{align}
From the viewpoint of $S^{d-1}$, the edge factor $Z_{\text{edge}}$ similarly packages an infinite tower of $\mathfrak{so}(d)$ irreps. Our strategy is to apply the branching rule $\mathfrak{so}(d{+}2)\to \mathfrak{u}(1)\oplus \mathfrak{so}(d)$ \cite{tsukamoto_spectra_1981} to \eqref{eq:Zprebranch},
\begin{align}
    Z^{\text{1-loop}}_{\text{PI}}\!\left[S^{d+1}\right] \sim \prod \bigl(\mathfrak{u}(1)\oplus \mathfrak{so}(d)\text{ irreps}\bigr)\;,
\end{align}
which makes the $\mathfrak{so}(d)$ content explicit.

Carrying out this analysis, explained in detail in the companion work \cite{Law:2025ktz}, we obtain a refined form of \eqref{eq:Zedgeold}, valid for any $d\geq 3$:
\begin{empheq}[box=\fbox]{align}\label{eq:Zedgegravresult}
     Z_{\rm edge }& \propto \det\nolimits'\!\left| -\nabla_0^2-\tfrac{d-1}{\ell^2_\text{dS}}   \right|  \nn\\
      & \quad \times \det\nolimits'_{-1}\!\left| -\nabla_1^2-\tfrac{d-2}{\ell^2_\text{dS}} \right|^{\frac12}
      \times \det\nolimits'\!\left(-\nabla_0^2 \right)^{\frac12} \; .
\end{empheq}
Here $-\nabla_0^2$ and $-\nabla_1^2$ denote the scalar and transverse-vector Laplacians on $S^{d-1}$ of radius $\ell_\text{dS}$. Primes indicate omission of zero modes, while the subscript ``$-1$’’ denotes that the harmonic product is extended to begin at degree $l=-1$. Expression \eqref{eq:Zedgegravresult} makes manifest the ghost-like contributions: a tachyonic vector, two tachyonic scalars, and a massless scalar.

\subsection{Nonlinear realization of $SO(d+2)$}

The decomposition \eqref{eq:Zedgegravresult} reveals a nontrivial pattern of shift symmetries,
suggesting an underlying nonlinear realization of the de Sitter isometry group. Specifically, the (inverse of) determinants in \eqref{eq:Zedgegravresult} are uniquely reproduced by the quadratic actions
\begin{align}
	S\left[A \right]  \propto&\,  
    \int_{S^{d-1}} \frac{1}{4} F_{\mu\nu}F^{\mu\nu} -\frac{d-2}{\ell^2_\text{dS}} A_\mu A^\mu,  \label{eq:Aedgeaction} \\
	S\left[\phi^a \right]  \propto&\,  
    \int_{S^{d-1}}   \partial_\mu \phi^a  \partial^\mu \phi_a-\frac{d-1}{\ell^2_\text{dS}}  \phi^a \phi_a, \label{eq:bendedgeaction}\\
	S\left[\chi \right]  \propto&\,  
    \int_{S^{d-1}}  \partial_\mu \chi \, \partial^\mu \chi \; .\label{eq:chiedgeaction}
\end{align}
Here $F_{\mu\nu}=\nabla_\mu A_\nu-\nabla_\nu A_\mu$, and the two tachyonic scalars are grouped into an $SO(2)$ vector $\phi^a$ ($a=1,2$) with metric $\delta_{ab}$. We suppress the volume form $\int_{S^{d-1}} d^{d-1}x \sqrt{g}$ to $\int_{S^{d-1}}$. For any $d\geq 3$, these actions exhibit abelian shift symmetries
\begin{align}\label{eq:shiftsym}
	\delta A_\mu = \xi^\text{KV}_{\mu} \; , \; \delta\phi^a = \nabla^\lambda\xi^\text{CKV,(a)}_{\lambda} \;, \; \delta \chi =\text{constant} \; ,
\end{align}
corresponding to the zero modes of the operators in \eqref{eq:Zedgegravresult}, where $\xi^\text{KV}_\mu$ and $\xi^\text{CKV}_\mu$ denote Killing and conformal Killing vectors on the round $S^{d-1}$. These symmetries realize the $\mathfrak{so}(d)$ decomposition of the $\mathfrak{so}(d+2)$ generators:
\Yvcentermath1
\begin{align}
    \underbrace{\yng(1,1)}_{\mathfrak{so}(d+2)}
\rightarrow
\underbrace{\yng(1,1)
\;\oplus\;
2\,\yng(1)
\;\oplus\;
  \bullet
}_{\mathfrak{so}(d)} \;,
\end{align}
with the $\mathfrak{so}(d)$ irreps on the right-hand side carried precisely by the shift symmetries \eqref{eq:shiftsym}

\subsection{Geometric fluctuations of the horizon}

It is natural to ask whether the explicitly determined spectra \eqref{eq:Aedgeaction}-\eqref{eq:chiedgeaction} admit a geometric interpretation. Indeed, the spectra and the shift symmetries \eqref{eq:shiftsym} they display are sufficiently constraining to infer a natural interpretation in terms of fluctuations of the Euclidean horizon $S^{d-1}\subset S^{d+1}$,
a maximally (totally geodesic) embedded surface that Lorentzian-continues to the cosmic horizon
(Fig.~\ref{pic:dSSP}).

The simplest fluctuations are transverse displacements of $S^{d-1}$, described by two scalar fields $\phi^a$. The induced metric on the fluctuated surface takes the form \cite{Clark:2005ht,Clark:2007rn,Goon:2011qf,Goon:2011uw,Burrage:2011bt}
\begin{align}\label{eq:inducedmetric}
    G_{\mu\nu}[\phi^a] = f^2 g_{\mu\nu} + B_{\mu\nu} \,,
\end{align}
where $g_{\mu\nu}$ is the round $S^{d-1}$ metric (radius $\ell_\text{dS}$), and \cite{Law:2025ktz}
\begin{align}
    f^2 = 1-\phi^a\phi_a \;, \quad 
    B_{\mu\nu} = \ell^2_\text{dS}\,\partial_\mu \phi^a \partial_\nu \phi_a + O(\phi^4)\,.
\end{align}
The worldvolume action 
\begin{align}\label{eq:bendaction}
  S[\phi^a] &\propto \int_{S^{d-1}} d^{d-1}x \,\sqrt{\det G[\phi^a]} \;,
\end{align}
expanded in small $\phi^a$, reproduces \eqref{eq:bendedgeaction}. Geometrically, the tachyonic mass $-(d-1)/\ell_\text{dS}^2$ reflects that the $S^{d-1}$ is a maximal embedding: constant transverse translations reduce its area. 

A second class of fluctuations corresponds to intrinsic diffeomorphisms generated by a vector field $A_\mu$. In fact, the quadratic action \eqref{eq:Aedgeaction} can be written as
\begin{gather}\label{eq:diffaction}
    S[A] \propto \int_{S^{d-1}} \; M^{\mu\nu}M_{\mu\nu} - \left(M\indices{^\lambda_\lambda}\right)^2 \;,
\end{gather}
in terms of the Lie derivative
\[
M_{\mu\nu} \equiv \mathcal{L}_A g_{\mu\nu} = \nabla_\mu A_\nu + \nabla_\nu A_\mu \, .
\]
of the round metric $g_{\mu\nu}$.

\begin{figure}[H]
	\centering
	\includegraphics[width=0.45\textwidth]{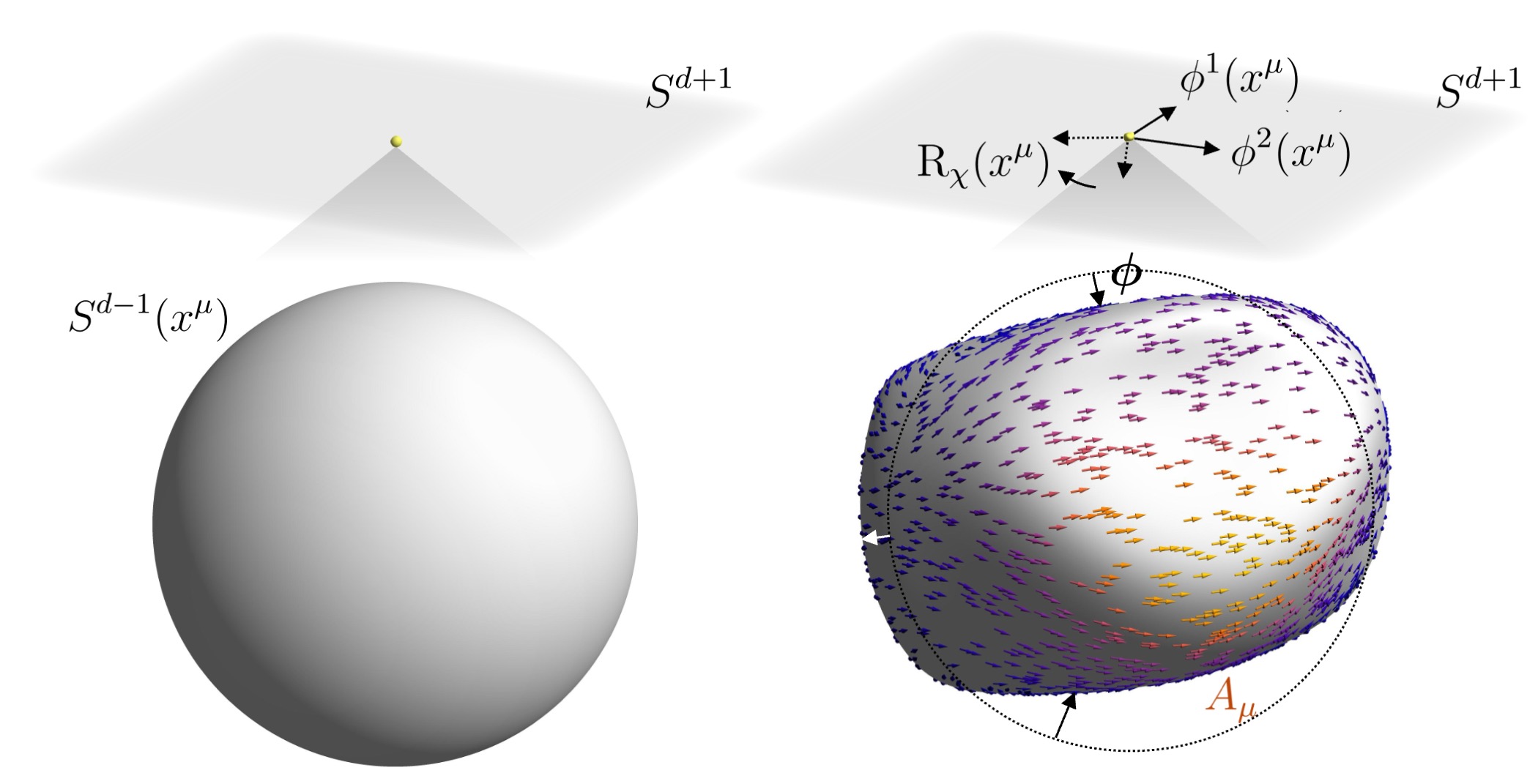}
	\caption{$Z_{\rm edge}$ of gravity encodes geometric fluctuations of the Euclidean horizon $S^{d-1}\subset S^{d+1}$: transverse bendings $\phi^a$, intrinsic diffeomorphisms $A_\mu$, and normal-bundle twists $\chi$.}
	\label{pic:embed}
\end{figure}

Finally, the $S^{d-1}$ carries a two-dimensional normal bundle inside $S^{d+1}$ with structure group $SO(2)$. The massless scalar $\chi$ can be interpreted as an angle field parameterizing a flat connection $\partial_\mu \chi$ on this bundle, with the minimal action \eqref{eq:chiedgeaction}. A summary of the three fluctuation types is shown in Fig.~\ref{pic:embed}.
We close with a brief remark. The determinants \eqref{eq:Zedgegravresult} can equally be obtained from quadratic actions for a Euclidean horizon embedded in {\it Lorentzian} de Sitter, $S^{d-1}\subset dS_{d+1}$. The actions \eqref{eq:Aedgeaction}–\eqref{eq:chiedgeaction} reappear with a relative minus sign between $\phi^1$ and $\phi^2$, and a careful evaluation of the $S^{d-1}$ path integrals yields the first factor in \eqref{eq:Zedgegravresult} with a dimension-dependent overall phase \cite{Law:2025ktz}.

%%%%%%%%%%%%%%%%%%%%%%%%%%%%%%%%%%%%%%%%%%%%%%%%%%%%%%%%%

\section{Symmetric tensor fields on $S^{d+1}$}

We now consider massless higher-spin gauge fields, central to various proposals for de Sitter holography \cite{Anninos:2011ui,Anninos:2012ft,Anninos:2013rza,Anninos:2017eib}. The quadratic action for a spin-$s$ gauge field is invariant under the gauge transformations \cite{Fronsdal:1978rb}
\begin{align}
    \delta \phi_{\mu_1 \cdots \mu_s} = \nabla_{(\mu_1} \xi_{\mu_2 \cdots \mu_s)} \; .
\end{align}
The corresponding one-loop $S^{d+1}$ partition functions \cite{Law:2020cpj} factorize as in \eqref{introeq:PIsplit}, with edge contributions appearing in a ``collapsed'' form in \cite{Anninos:2020hfj}, directly analogous to the unrefined graviton result \eqref{eq:Zedgeold}.

The branching rule method sketched in the last section can be applied directly to this case. Alternatively, following the graviton analysis, one may use shift symmetries as a guiding principle. For instance, the global gauge transformations of a massless spin-3 theory are generated by spin-2 Killing tensors, transforming in the two-row $(2,2)$ irrep of $\mathfrak{so}(d+2)$. For any $d \geq 4$ this decomposes as
\Yvcentermath1
\begin{align}
    & \qquad \underbrace{\yng(2,2)}_{\mathfrak{so}(d+2)} \nn\\
    &\longrightarrow
    \underbrace{\bullet
    \,\oplus\,
    2\,\yng(1)
    \,\oplus\,
    3\,\yng(2)
    \,\oplus\,
    \yng(1,1)
    \oplus
    2\,\yng(2,1)
    \oplus
    \yng(2,2)}_{\mathfrak{so}(d)} \, .
\end{align}
Summing over spin-$0$, $1$, and $2$ fields whose actions are invariant under shifts by the corresponding spherical harmonics on $S^{d-1}$ reproduces (the kinematic part of) the collapsed formula of \cite{Anninos:2020hfj}. This reasoning extends to arbitrary spin and all $d \geq 3$, further reinforcing the link between $Z_{\rm edge}$ and shift symmetries.
\begin{table*}[ht]
	%\centering
	\renewcommand{\arraystretch}{1.5} % Adjust row height for better spacing
    \caption{\label{tab:tablePM}%
Field contents of $Z_{\rm edge}$ for various classes of theories on $S^{d+1}$. For massless fields, totally symmetric and antisymmetric tensor gauge fields give rise to two distinct types of edge theories: shift-symmetric and gauge theories. }
    \begin{ruledtabular}
	\begin{tabular}{cccc}
		 & {$p$-form} \cite{Mukherjee:2023ihb,Ball:2024xhf} & {Rank-$s$ totally symmetric tensor} \cite{Law:2025ktz} & {Mixed-symmetry} \\ 
		\colrule
		{Massive} & {Massive $(p-1)$-form} & {Massive spin $\leq s-1$} & {??} \\ 
		{Partially massless} \cite{Deser:1983tm,DESER1984396,Higuchi:1986py,Brink:2000ag,Deser:2001pe,Deser:2001us,Deser:2001wx,Deser:2001xr,Zinoviev:2001dt,Hinterbichler:2016fgl} & {(Not applicable)} & {Shift-symmetric + massive spin $\leq s-1$} & {??} \\ 
		{Massless} & {Massless $(p-1)$-form} & {Shift-symmetric spin $\leq s-1$} & {(Not applicable)} \\ 
	\end{tabular}
    \end{ruledtabular}
\end{table*}
In table \ref{tab:tablePM}, we summarize the field contents of $Z_{\rm edge}$ for a range of massive and (partially) massless theories. With the $\mathfrak{so}(d+2)\to \mathfrak{u}(1)\oplus \mathfrak{so}(d)$ branching rule method sketched above and developed in detail in \cite{Law:2025ktz}, a natural next step is to extend this summary to more general mixed-symmetry tensor fields in arbitrary dimensions.

%%%%%%%%%%%%%%%%%%%%%%%%%%%%%%%%%%%%%%%%%%%%%%%%%%%%%%%%%

\section{Gravitons on $S^2\times S^{d-1}$}

%\VL{Say early that this is a nontrivial cross-saddle test for the story we uncovered in graviton case on dS and reiterate in last paragraph on Intro} 
We now examine whether the shift-symmetry structure of the gravitational edge sector persists in a subleading saddle of \eqref{introeq:Z}, namely $S^2 \times S^{d-1}$. The Lorentzian version of this geometry is the static Nariai spacetime \cite{1950SRToh..34..160N}, the extremal limit of Schwarzschild–de Sitter, where the black hole and cosmological horizons coincide at radius $r_N = \sqrt{\frac{(d-1)(d-2)}{2\Lambda}}$. The spacetime factorizes as $dS_2 \times S^{d-1}$ with $dS_2$ and $S^{d-1}$ radii $\ell_N = \frac{r_N}{\sqrt{d-2}}$ and $r_N$ respectively, and Wick rotation maps the two horizons to the poles of the $S^2$ factor, yielding $S^2 \times S^{d-1}$.

In a companion work \cite{Law:2025yec} (see also \cite{Mukherjee:2025xlt}), we computed the complete 1-loop graviton partition function for any $d \geq 3$. The result is the analog of \eqref{introeq:PIsplit}, \eqref{introeq:Zbulk}, and \eqref{eq:Zedgegravresult}:
\begin{align}\label{eq:ZPINariai}
    Z^\text{1-loop}_\text{PI}\!\left[S^2 \times S^{d-1}\right]
    = Z_\text{bulk}(\beta=\beta_N)\, Z_\text{edge} \;,
\end{align}
with $Z_\text{bulk}$ the thermal partition function of an ideal graviton gas in the Lorentzian Nariai background at inverse temperature $\beta_N = 2\pi \ell_N$. Equation~\eqref{introeq:charQNM} is replaced by the corresponding Nariai character, summarized in the Appendix. The edge factor in \eqref{eq:ZPINariai} is
\begin{empheq}[box=\fbox]{align}\label{eq:ZedgegravNariai}
     Z_{\rm edge } \;\propto\; 
     \left(
        \det\nolimits'_{-1}\!\left| -\nabla_1^2-\tfrac{d-2}{r_N^{2}} \right|^{\frac12}
        \det\nolimits'\!\left(-\nabla_0^2\right)^{\frac32}
     \right)^{2},
\end{empheq}
receiving ghost-like contributions from two identical sets of fields on $S^{d-1}$ (radius $r_N$), each a tachyonic vector plus three massless scalars. As before, $-\nabla_0^2$ and $-\nabla_1^2$ are the scalar and transverse-vector Laplacians, primes indicate omission of zero modes, and the subscript ``$-1$'' extends the spherical-harmonic product to $l=-1$. The doubling naturally reflects the two Nariai horizons of equal radius.

Examining one copy, the effective masses differ from the pure dS case \eqref{eq:Zedgegravresult}, showing that the {\it dynamics} of graviton edge modes is sensitive to geometry beyond the intrinsic $S^{d-1}$. This contrasts with $p$-forms, whose spectra are insensitive to such details \cite{Ball:2024hqe,Ball:2024xhf}. Table~\ref{tab:tableBH} compares the edge spectra on $S^{d+1}$ and $S^2\times S^{d-1}$.
\begin{table*}[ht]
\renewcommand{\arraystretch}{1.5} % Adjust row height for better spacing
\caption{\label{tab:tableBH}%
Field content of $Z_\text{edge}$ for gravitons on $S^{d+1}$ and $S^2\times S^{d-1}$. Here $k$ labels the shift-symmetry level \cite{Bonifacio:2018zex,Bonifacio:2019hrj}. The last line shows the contribution of each edge field type to the 4D logarithmic coefficient, defined by $\log Z_{\rm edge}|_\text{log-div.}=\alpha_\text{edge}\log(\mu \ell)$, with $\mu$ the UV mass scale and $\ell$ the $S^{d-1}$ radius. %\VL{$l_{dS}$ not for Nariai, just say $l$}
}
\begin{ruledtabular}
\begin{tabular}{c|cc}
& Round $S^{d+1}$ ($d\geq 3$) & $S^2\times S^{d-1}$ ($d\geq 3$) \\
\colrule
Isometry & $SO(d+2)$ & $SO(3)\times SO(d)$ \\
Lorentzian geometry & $dS_{d+1}$ static patch & Static Nariai \\
$Z_\text{edge}$ content & \thead{one $k=0$ vector + two $k=1$ scalars \\+ one $k=0$ scalar} & $(\text{one $k=0$ vector + three $k=0$ scalars})\times 2$ \\
4D log-coefficient & $\alpha^{S^4}_\text{edge} = -\tfrac{16}{3} = -\tfrac13 - \tfrac73\times 2 - \tfrac13$ & $\alpha^{S^2\times S^2}_\text{edge} = -\tfrac{8}{3} = \left(-\tfrac13 - \tfrac13 \times 3\right)\times 2$
\end{tabular}
\end{ruledtabular}
\end{table*}

\subsection{Shift symmetries and the geometric interpretation}

In a single copy, the tachyonic vector has the same mass (in units of the round $S^{d-1}$ radius) as in the $S^{d+1}$ case, while the two tachyonic scalars of \eqref{eq:Zedgegravresult} are replaced by two massless scalars. This spectrum reflects that $Z_\text{edge}$ again nonlinearly realizes the background isometries: the number of shift symmetries (the $SO(d)$ generators from the vector plus three constant shifts) equals the dimension of the isometry group $SO(3)\times SO(d)$.

In terms of geometric fluctuations of the horizon $S^{d-1}$, the vector has the same mass (in units of $r_N$) as in the round $S^{d+1}$ case and continues to describe diffeomorphisms intrinsic to $S^{d-1}$, while another massless scalar continues to encode the $SO(2)$ normal-bundle connection, i.e. local rotations of the two transverse directions. 

The remaining question is whether the two other massless scalars can be viewed as coordinates transverse to the Euclidean horizon. Indeed, embedding a round $S^{d-1}$ into $S^2\times S^{d-1}$ gives the induced metric \eqref{eq:inducedmetric}, but with 
\begin{align}
    f^2 = 1 \;, \qquad 
    B_{\mu\nu} = \ell_N^2\, \partial_\mu \phi^a \partial_\nu \phi_a + O(\phi^4) \;.
\end{align}
The trivial dependence of $f^2$ on $\phi^a$ makes these scalars massless: constant shifts of the $S^{d-1}$ along the $S^2$ factor leaves its radius unchanged.

%%%%%%%%%%%%%%%%%%%%%%%%%%%%%%%%%%%%%%%%%%%%%%%%%%%%%%%%%

\section{Discussion}

We have uncovered rich codimension-2 structures in one-loop partition functions around the $S^{d+1}$ and $S^2 \times S^{d-1}$ saddles of the Euclidean gravitational path integral \eqref{introeq:Z}. For gravity and higher-spin gauge fields, these structures exhibit novel shift symmetries. In the gravitational case, the resulting spectra admit a geometric interpretation as fluctuations of the de Sitter or Nariai horizons. Below we highlight several broader connections, with further discussion in \cite{Law:2025ktz,Law:2025yec}.

\subsection{Cornering Gravitational Edge Modes}

A natural question is whether these explicit edge spectra admit a direct Lorentzian, Hamiltonian description. In the gravity case, there has been substantial progress in understanding the classical phase space of gravity in a finite region. It is found that the codimension-2 corner $S$ of any causal diamond has an extended corner symmetry group \cite{Donnelly:2016auv,Speranza:2017gxd,Freidel:2020xyx,Freidel:2020svx,Freidel:2020ayo,Ciambelli:2022vot}
\begin{align}\label{eq:ECS}
\mathrm{Diff}(S) \ltimes SL(2,\mathbb{R})^S \ltimes \big(\mathbb{R}^2\big)^S \; ,
\end{align}
where $\mathrm{Diff}(S)$ are diffeomorphisms intrinsic to $S$, $SL(2,\mathbb{R})^S$ acts on its normal bundle, and $\big(\mathbb{R}^2\big)^S$ represent corner translations. In both the $dS_{d+1}$ static patch and the Nariai black hole, the relevant corners are the cosmic and black hole horizons, each with topology $S^{d-1}$.

Our gravitational edge spectra (Table \ref{tab:tableBH}) can be viewed as the Euclidean realization of \eqref{eq:ECS}. In the static patch, the vector $A_\mu$ parametrizes surface diffeomorphisms $\mathrm{Diff}(S^{d-1})$, the tachyonic scalars $\phi^a$ the translations $\big(\mathbb{R}^2\big)^{S^{d-1}}$, and $\chi$ the Euclidean analogue $U(1)^{S^{d-1}}$ of the boost subgroup $SO(1,1)^{S^{d-1}} \subset SL(2,\mathbb{R})^{S^{d-1}}$. The Nariai case admits an analogous mapping.

\subsection{Edge modes as Goldstones?}

For gravity and higher-spin gauge theories on $S^{d+1}$, the relation between $Z_\text{edge}$ and shift symmetries suggests that edge modes may be understood as Goldstone-like excitations of a symmetry breaking mechanism.

From the $dS_{d+1}$ perspective, specifying a static patch restricts the full diffeomorphism group to the subgroup that preserves it. The split \eqref{introeq:PIsplit} can then be seen as the Euclidean imprint of this mechanism. Defining $Z_\text{bulk}$ presumably involves selecting a reference $S^{d-1}$, the Euclidean horizon, which reduces the diffeomorphisms to those leaving this surface fixed, while $Z_\text{edge}$ acts as a compensating dressing that restores invariance of the full partition function $Z_\text{PI}$. One may further speculate that the edge modes represent diffeomorphisms that deform the Euclidean horizon, reflecting the degrees of freedom of a quantum reference frame \cite{Carrozza:2021gju,Carrozza:2022xut,Jensen:2023yxy,DeVuyst:2024khu,AliAhmad:2024wja,AliAhmad:2024vdw,Kirklin:2024gyl,DeVuyst:2024fxc, fewster2025quantum, fewster2025semi}, and can be gauged away once a physical observer, clock, or frame is introduced \cite{Anninos:2011af,Goeller:2022rsx,Chandrasekaran:2022cip,Chen:2024rpx,Maldacena:2024spf}. A natural question is whether this perspective can be developed into a systematic derivation of the factorization \eqref{introeq:PIsplit}, ideally in a form extendable to interacting theories{, where one might not expect a clean bulk-edge factorization of the $S^{d+1}$ path integral}.

\subsection{Revisiting the Gibbons-Hawking prescription}

The appearance of $Z_\text{edge}$ motivates reexamining the heuristic argument that $S^{d+1}$ path integrals compute thermal traces in a $dS_{d+1}$ static patch. The subtleties lie precisely at the codimension-2 fixed point sets (Fig.~\ref{pic:dSSP}): the Lorentzian bifurcation surface and the Euclidean horizon where the $U(1)$ thermal circle degenerates.

Beyond one loop, the trace interpretation of the full gravitational path integral \eqref{introeq:Z} is even less clear. It has never been derived from a Hamiltonian framework, and the conformal factor problem—responsible for the one-loop phase \cite{Gibbons:1978ac,Polchinski:1988ua,Law:2020cpj,Maldacena:2024spf,Shi:2025amq,Ivo:2025yek,Law:2025yec}—has led to renewed scrutiny of the Euclidean configuration-space path integral \cite{Banihashemi:2024weu,Horowitz:2025zpx} and proposals of Lorentzian formulations \cite{Marolf:2022ybi}. Another approach to clarifying the thermal interpretation of \eqref{introeq:Z} is to introduce a finite boundary \cite{Banihashemi:2022jys,Anninos:2024wpy,Silverstein:2024xnr}. These directions may ultimately be compatible with the no-boundary Euclidean path integral studied here, and extending them to one loop would provide a sharper basis for comparison.

\subsection{More General Spacetimes}

Beyond $S^{d+1}$ and $S^2\times S^{d-1}$, the path integral \eqref{introeq:Z} admits other saddles. In $d+1=4$, see \cite{Anninos:2025ltd} for a recent discussion. Notable examples include the Page manifold \cite{Page:1978vqj} with topology $\mathbb{CP}^2 \# \overline{\mathbb{CP}}^2$ and its higher-dimensional generalizations \cite{Gibbons:2004uw}, corresponding to Euclidean continuations of rotating Nariai black holes with special complexified parameters. For odd $d+1$, further saddles arise as smooth quotients of the round $S^{d+1}$. A canonical case is the Lens spaces $L(p,q)$, the $\mathbb{Z}_p$ quotients of $S^3$, interpretable as grand canonical partition functions in a $dS_3$ static patch with chemical potential \cite{Castro:2011xb}. Analogues of \eqref{introeq:PIsplit}, \eqref{introeq:Zbulk}, and \eqref{eq:Zedgegravresult} for massive higher-spin fields on $L(p,q)$ were obtained in \cite{Law:2021hwc}.

Beyond the $\Lambda>0$ setting, formulas such as \eqref{introeq:PIsplit} and \eqref{eq:ZPINariai} extend to static black hole spacetimes \cite{Denef:2009kn,Law:2022zdq}. Here $Z_\text{bulk}$ follows from the black hole versions of \eqref{introeq:Zbulk} and \eqref{introeq:charQNM}, while $Z_\text{edge}$, present only for spin $s \geq 1$, captures QNMs that do not Wick-rotate to Euclidean modes with low Matsubara frequencies \cite{Castro:2017mfj,Keeler:2018lza,Keeler:2019wsx,Grewal:2022hlo}. For rotating BTZ black holes, $Z_\text{edge}$ is essential for reproducing the non-local $T^{3/2}$ correction to the one-loop graviton partition function in the low-temperature limit \cite{Kapec:2024zdj}.

%%%%%%%%%%%%%%%%%%%%%%%%%%%%%%%%%%%%%%%%%%%%%%%%%%%%%%%%%

\begin{acknowledgments}

\section{Acknowledgments}
It is a great pleasure to thank Douglas Stanford for stimulating conversations, and especially Dionysios Anninos, Batoul Banihashemi, Frederik Denef and Laurent Freidel for useful discussions and comments on the draft. 
AL  was supported in part by the Stanford Science Fellowship, a Simons Investigator award, and NSF Grant PHY-2310429. VL was supported in part by the U.S. Department of Energy grant DE-SC0011941.

\end{acknowledgments}

\appendix

\section{Gravitons on $S^{d+1}$}

The one-loop path integral for gravitons on the round $S^{d+1}$, of radius 
$\ell_\text{dS} \equiv \sqrt{\tfrac{d(d-1)}{2\Lambda}}$, can be written in terms of functional determinants \cite{Gibbons:1978ji, Christensen:1979iy, Fradkin:1983mq, Allen:1983dg, Taylor:1989ua, GRIFFIN1989295, Mazur:1989ch, Vassilevich:1992rk, Volkov:2000ih, Polchinski:1988ua, Anninos:2020hfj, Law:2020cpj}:
\begin{align}
      Z^\text{1-loop}_\text{PI}\!\left[S^{d+1} \right] 
      =  \frac{i^{\,d+3}}{\text{Vol}\left(G\right)_{\text{PI}}}\,
      \frac{\det\nolimits_{-1}'\!\left|-\nabla_{1,S}^2-\tfrac{d}{\ell_\text{dS}^2}\right|^{1/2}}
           {\det_{-1}\!\left|-\nabla_{2,S}^2+\tfrac{2}{\ell_\text{dS}^2}\right|^{1/2}} \;.
\end{align}
Here $-\nabla_{1,S}^2$ and $-\nabla_{2,S}^2$ denote the transverse vector Laplacian and the symmetric transverse–traceless (STT) spin-2 Laplacian on $S^{d+1}$. A prime indicates omission of zero modes of the corresponding Laplace-type operator, and the subscript “$-1$’’ denotes that the product over spherical harmonics is extended to start at the $SO(d{+}2)$ label $L=-1$. The isometry group is $G=SO(d+2)$, whose volume is
\begin{align}\label{eq:ZPIvolgrav}
	\text{Vol}(G)_{\rm PI} = \text{Vol}(G)_c \left( \frac{\text{Vol}(S^{d-1})\,\ell_\text{dS}^{\,d-1}}{8\pi G_N\, d(d+2)}\right)^{\tfrac{\dim SO(d+2)}{2}} \;,
\end{align}
measured in the $G_N$-dependent metric induced by the path-integral measure. The ``canonical" volume of $SO(d+2)$ is defined using the invariant group metric normalized so that the minimal $SO(2)$ orbits have length $2\pi$:
\begin{align}\label{eq:volumes}
	\text{Vol}(G)_c = \prod_{k=2}^{d+2} \text{Vol}\left(S^{k-1}\right)\;, 
	\;
	\text{Vol}\left(S^n\right) = \frac{2\pi^{\tfrac{n+1}{2}}}{\Gamma\!\left(\tfrac{n+1}{2}\right)} \;.
\end{align}

After some manipulations \cite{Anninos:2020hfj}, one finds the factorization \eqref{introeq:PIsplit}. 
In the bulk sector \eqref{introeq:Zbulk}, the Harish–Chandra character for a graviton takes the form
\begin{align}
	&\quad \chi(t)	\nn\\
    &=  \sum_{I=0}^2 \sum_{n=0}^\infty \sum_{l=2}^\infty 
    D^d_{l,I} \left(e^{-\frac{I+2n+l}{\ell_\text{dS}}|t|} + e^{-\frac{d-I+2n+l}{\ell_\text{dS}}|t|}\right)\nn\\
    &=  \left[ \frac{(d+2)(d-1)}{2}\, \frac{q^d+1}{|1-q|^d} 
       - d\,\frac{q^{d+1}+q^{-1}}{|1-q|^d}\right]_+ \;,
\end{align}
where $q\equiv e^{-t/\ell_\text{dS}}$. 
On the first line, the exponents encode the physical QNM frequencies for a graviton in the $dS_{d+1}$ static patch \cite{Lopez-Ortega:2006aal}. 
Here $I=0,1,2$ label scalar, vector, and tensor modes, respectively; $n$ is the overtone number; and $l$ is the total $SO(d)$ angular momentum. 
The degeneracy $D^d_{l,I}$ is the dimension of a two-row representation of $SO(d)$ \cite{rubin1984eigenvalues,Higuchi:1986wu}. 
The notation $[\,\cdots\,]_+$ in the second line denotes the following ``flipping'' operation on a small-$q$ expansion
\begin{align}\label{eq:flipnotation}
	\left[\sum_k c_k q^k \right]_+  
    \equiv \sum_k c_k q^k - c_0 - \sum_{k<0} c_k \left(q^k + q^{-k}\right) \;. 
\end{align}

For completeness, the full expression for the edge contribution \eqref{eq:Zedgeold} is
\begin{align}
     Z_{\rm edge}	
    & =\frac{i^{d+3} }{ \text{Vol}(G)_c } 
         \left(\frac{32\pi^3 G_N}{\text{Vol}(S^{d-1})\ell_\text{dS}^{d-1}}\right)^{\frac{\dim SO(d+2)}{2}} \nn\\
    &\quad \times \exp\bigg\{-\int_0^\infty \frac{dt}{2t}\frac{1+q}{1-q}\nn\\
   & \qquad  \times \left[(d+2)\frac{q^{d-1}+q^{-1}}{(1-q)^{d-2}}-\frac{q^{d}+q^{-2}}{(1-q)^{d-2}} \right]_+ \bigg\}\;.
\end{align}

\section{Gravitons on $S^2\times S^{d-1}$}

The one-loop path integral for gravitons on $S^2\times S^{d-1}$ ($d\geq 3$), with radii 
$\ell_N \equiv \sqrt{\tfrac{d-1}{2\Lambda}}$ and $r_N \equiv \sqrt{d-2}\,\ell_N$, can be written in terms of functional determinants \cite{Volkov:2000ih,Law:2025yec}:
\begin{align}
     & \quad Z^\text{1-loop}_\text{PI}\!\left[S^2 \times S^{d-1} \right] \nn\\
 &    =  \frac{1}{\text{Vol}\!\left(G\right)_{\text{PI}}}\,
       \left( \frac{d-1}{4\Lambda}\right)^{\frac12}
       \frac{\det\nolimits'\!\left( -\nabla_{1,N}^2-\tfrac{d-2}{r_N^2}\right)^{\frac12}}
            {\det\!\left| D^2\right|^{\frac12}} \;.
\end{align}
Here $-\nabla_{1,N}^2$ is the transverse vector Laplacian on $S^2\times S^{d-1}$, while 
\begin{align}\label{eq:TTLaplacian}
    D^2_{\mu\nu\alpha\beta}  \equiv - g_{\mu\alpha}g_{\nu\beta}\nabla^2  - 2 R_{\mu \alpha \nu \beta} 
\end{align}
is the kinetic operator acting on the transverse–traceless (TT) part of the graviton. 
The isometry group is $G=SO(3)\times SO(d)$, whose path-integral volume is related to the canonical volume,
\begin{align}
    \text{Vol}(G)_c = 8\pi^2 \prod_{I=1}^{d-1} \text{Vol}\!\left(S^I\right) \;,
\end{align}
through
\begin{align}\label{eq:NariaiGPI}
	& \quad \text{Vol}\!\left(G\right)_{\text{PI}} \nn\\
	& = \left( \frac{A}{8\pi G_N}\,\frac{\ell_N^4}{3} \right)^{\frac32}
	  \left( \frac{A}{8\pi G_N}\,\frac{r_N^2\ell_N^2}{d}   \right)^{\tfrac{d(d-1)}{4}}
	  \text{Vol}\!\left(G\right)_{c} \,,
\end{align}
where $A=2\times r_N^{\,d-1}\,\text{Vol}\!\left(S^{d-1}\right)$ is the total area of the black-hole and cosmological horizons of the Nariai spacetime.

One finds the factorization \eqref{eq:ZPINariai}. The bulk part takes the same form as \eqref{introeq:Zbulk}, but with
\begin{align}
    \chi(t) &\equiv \sum_{I=0}^2\sum_{l=2}^\infty \sum_{n=0}^\infty D^d_{l,I} \, \left( e^{-i \omega^I_{nl}|t| } +e^{-i {\tilde\omega}^I_{nl}|t| }\right)\nn\\
    &= \sum_{I=0}^2\sum_{l=2}^\infty D^d_{l,I} \, \frac{e^{-\Delta_{I,l}\frac{|t|}{\ell_N}}+e^{-\bar\Delta_{I,l}\frac{|t|}{\ell_N}}}{1-e^{-\frac{|t|}{\ell_N}}} \;. 
\end{align}
On the first line, the exponents encode the graviton QNM spectrum in the Lorentzian Nariai geometry \cite{Vanzo:2004fy,Lopez-Ortega:2007llk,Lopez-Ortega:2009flo,Venancio:2020ttw}:
\begin{gather}
    i \omega^I_{nl} = \frac{\Delta_{I,l}+n}{\ell_N} \qquad \text{and} \qquad i {\tilde\omega}^I_{nl} =\frac{\bar\Delta_{I,l}+n}{\ell_N}, \nn\\
    \text{where}\;\; \Delta_{I,l} =1-\bar\Delta_{I,l}= \frac12 + i \nu_{I,l} \;\; \text{with} \nn\\
      \nu_{I,l}\equiv  \sqrt{\frac{l\,(l+d-2)-(2-I)(d-2+I)}{d-2} -\frac{1}{4} } \;,
\end{gather}
with $I=0,1,2$ labeling scalar, vector, and tensor modes, $n$ the overtone number, $l$ the total $SO(d)$ angular momentum, and $D^d_{l,I}$ the degeneracies.

On the other hand, the edge partition function is
\begin{align}\label{eq:Zedgedet}
    &\quad Z_{\rm edge} \nn\\
    &= \frac{1}{\text{Vol}\!\left(G\right)_{\text{PI}}} 
       \left[ \det\nolimits_{-1}' \!\left| -\nabla_1^2-\tfrac{d-2}{r^2_N}\right|^{\frac12}
              \det\nolimits' \!\left(-\nabla_0^2\right)^{\frac32} \right]^2 .
\end{align}

\bibliography{ref}% Produces the bibliography via BibTeX.

\end{document}